# Coronavirus and oil price crash

Claudiu Tiberiu ALBULESCU[1,2]*

[1] Management Department, Politehnica University of Timisoara, 2, P-ta. Victoriei, 300006 Timisoara, Romania.

[2] CRIEF, University of Poitiers, 2, Rue Jean Carbonnier, Bât. A1 (BP 623), 86022, Poitiers, France.

**Abstract**

Coronavirus (COVID-19) creates fear and uncertainty, hitting the global economy and amplifying the financial markets volatility. The oil price reaction to COVID-19 was gradually accommodated until March 09, 2020, when, 49 days after the release of the first coronavirus monitoring report by the World Health Organization (WHO), Saudi Arabia floods the market with oil. As a result, international prices drop with more than 20% in one single day. Against this background, the purpose of this paper is to investigate the impact of COVID-19 numbers on crude oil prices, while controlling for the impact of financial volatility and the United States (US) economic policy uncertainty. Our ARDL estimation shows that the COVID-19 daily reported cases of new infections have a marginal negative impact on the crude oil prices in the long run. Nevertheless, by amplifying the financial markets volatility, COVID-19 also has an indirect effect on the recent dynamics of crude oil prices.

* Corresponding author. E-mail address: claudiu.albulescu@upt.ro.

## 1. Introduction

The coronavirus (COVID-19) pandemic outbreaks Wuhan (Hubei region from China), where the first infection case is reported on December 31, 2019. 49 days after the release of the first coronavirus monitoring report by the World Health Organization (WHO) on January 21, 2020, over 100,000 people from more than 100 countries around the world are affected. Although COVID-19 does not present similar patterns in terms of fatality rate as compared to the 2002-2003 Severe Acute Respiratory Syndrome (SARS), or in terms of global spread as compared to the Spanish Flu pandemic of 1919, the new coronavirus is very contagious and triggers a lot of uncertainty in the real economy and financial markets.[1] COVID-19 negatively affects the overall demand, creating short-run volatility in the food prices,[2] and impending the mobility of workers and tourists. In addition, COVID-19 creates fear and additional stress on financial markets, where the price volatility is continuously increasing. Anticipating a strong decrease in the global demand in the next period, Saudi Arabia starts an oil price war on March 09, 2020, and floods the market with oil. In one single day, the crude oil price plunges with more than 20%. This shock spills over financial markets that crash during the same day (the Black Monday).

In this context, the present research tempts to investigate how the COVID-19 new infection cases affect the oil price, while controlling for the role of financial stress and volatility (VIX index), and the United States (US) economic policy uncertainty (EPU). To this end, we resort to an Autoregressive Distributed Lag (ARDL) specification, which allow us to see if the relationship between oil price, financial volatility, economic policy uncertainty and COVID-19 converge toward a long-run equilibrium, in the presence of both stationary and non-stationary series.

The relationship between stock prices and oil prices was intensively debated in the literature given the financialization of commodity markets (for a recent review, see Balcilar et al., 2019; Wen et al., 2019). However, the interaction between financial volatility on the one hand and oil prices on the other hand, has received less attention. Illing and Liu (2006) notice that oil price shocks correspond to picks in the financial stress index for Canada, and formally test this relationship. In fact, financial stress episodes affect the stock market returns and create volatility, whereas the stock prices and oil prices are highly correlated. Recent works in this line document a bidirectional relationship between oil prices and financial volatility (e.g. Basher and Sadorsky, 2016; Das et al., 2018; Nazlioglu et al. (2015).

---
[1] The fatality (death) ratio at global level was 11% for SARS (and less than 4% for COVID-19), whereas he Spanish Flu pandemic affected hundreds of millions of people.
[2] Supermarkets were emptied in several regions from Italy, Germany and the United Kingdom.

Another strand of literature investigates the relationship between the oil price and the economic policy uncertainty (EPU), resorting to the EPU index constructed by Baker et al. (2016). This relationship is particularly important for the US, the dollar being the transaction currency on oil markets. Oil prices influence the forecasts of macroeconomic variables and therefore the US EPU. At the same time, policy-induced uncertainty influences the asset prices in general, including oil prices. Therefore, recent papers investigate the both sides of the coin, looking to the reverse causality between the oil price and the EPU, and reporting mixed findings. A first set of papers (e.g. Antonakakis et al., 2014; Kang et al., 2017) shows that oil price shocks impact the US EPU. A second set of studies underlines, on contrary, that US EPU dynamics triggers movements in international oil prices (Aloui et al., 2016; Yang, 2019).

Finally, a separate strand of the literature assesses the interaction between both economic uncertainty and financial volatility, and the oil price. In a quantile regression framework, Reboredo and Uddin (2016) test to what extent the financial stress and policy uncertainty impact the energy and metal commodity prices in the US. They report a nonlinear effect of EPU and VIX on oil prices. With a focus on the shocks in oil prices, Degiannakis et al. (2018) document a time-varying effect on the US EPU and financial uncertainty, in specific periods.[3] However, none of these studies investigates the uncertainty, fear and panic triggered by COVID-19 on oil prices.

Therefore, our first contribution to the exiting literature is the assessment of the impact of COVID-19 on oil prices, while controlling for the role of US EPU and VIX. As far as we know, this is the first paper approaching this subject, which will certainly modify the public agendas for the following months.[4] We use the WHO official announcements in terms of COVID-19 new infection cases, and the West Texas Intermediate (WTI) for crude oil prices. Second, resorting to daily data, we look to both the short- and long-run relationship between oil prices, VIX and the US EPU. Our series are either I(1) or I(0), which recommends the use of the ARDL approach proposed by Pesaran et al. (2001). Third, we assess the effect of COVID-19 on oil prices, looking to the impact of the total number of new infection cases, as well as to the new cases reported in China and outside China. Finally, we test the robustness of our findings considering the BRENT crude prices.

---

[3] We notice that the relationship between economic uncertainty, financial volatility and oil prices is characterized by reverse causality. In fact, even the economic policy uncertainty represents a reliable cause of financial markets volatility as shown by Antonakakis et al. (2013), Li et al. (2019), Mei et al. (2018), Phan et al. (2020), Su et al., (2019), Tiwari et al. (2019), Zhenghui and Junhao (2019).
[4] Albulescu (2020) already investigated the effect of COVID-19 numbers on the financial markets volatility and reported a positive and significant impact.

The rest of the paper presents the literature (Section 2, data and methodology (Section 3), the empirical findings (Section 4) and the robustness results (Section 5). Finally, we conclude.

**2. Literature review**

Our paper stands at the crossroad of two strands of the literature, the first one investigating the relationship between the oil price and the financial volatility, and the second one addressing the interaction between the oil price and the economic policy uncertainty.

Within the first line of research, Illing and Liu (2006) investigate the financial instability in Canada and notice a positive correlation between the level of financial stress and oil price shocks. More recently, Gkillas et al. (2020) discover that financial stress helps to improve the oil price forecasts. Conducted an analysis over the period January 4, 2000 – May 26, 2017, the authors resort to different financial stress indexes and underline that particular attention should be paid to the financial volatility originated from the US, in order to correctly anticipate the oil price dynamics.

The implied financial volatility of S&P 500 (VIX) is often used as a proxy for financial stress and volatility. For example, Nazlioglu et al. (2015) conduct a volatility spillover analysis on the US for the period 1991 to 2014 and document a bidirectional relationship between VIX and the US EPU. The authors also report the dominance of the long-run volatility spillover. In the same spirit, Das et al. (2018) examine the dependence between stock prices, commodity prices, and financial stress. The authors resort to a nonparametric causality-in-quantile technique and focus on the US economy. They document a bilateral causality between oil prices and financial volatility. The result confirms the findings by Basher and Sadorsky (2016) who investigate the hedging property of oil prices and VIX, for emerging markets stock prices.

The relationship between EPU and oil prices reveals its turn the existence of a reverse causality situation. For example, Chen et al. (2020) apply a discrete wavelet transform and show that the impact of oil price shocks on EPU is positive at all frequencies, a result in line with Kang et al. (2017) and the findings reported by Antonakakis et al. (2014). Likewise, Kang et al. (2017) investigate the interaction between the US EPU and different economic and financial variables, including the oil price. They find that the oil supply shocks originated from US and non-US explain over 20% from the US EPU variation. Similar findings are reported by Antonakakis et al. (2014) who resort to the spillover index proposed by Diebold and Yilmaz (2012) and report that EPU shocks negatively respond to oil price shocks and *vice-versa*. Hailemariam et al. (2019) extend the analysis at the level of G7 countries using monthly data from 1997 to 2018. They underline the existence of a time-varying effect of oil prices on EPU.

Other studies highlight the impact of uncertainty on oil price returns. With a focus on extreme dependencies and resorting to a copula approach, Aloui et al. (2016) show that an increased EPU index has a positive effect on oil prices in periods that precede financial crises' outburst. A nonparametric causality-in-quantiles method is proposed by Shahzad et al. (2017) to investigate the causal effect of investors' sentiments and EPU on commodity prices, including oil. At the same time, Ma et al. (2018) resort to threshold forecast frameworks and notice that EPU is important to explain oil futures price. Similar, Yang (2019) performs both a causality and a spillover analysis between EPU and oil price shocks and emphasizes that oil prices behave as net receivers of information from EPU. Albulescu et al. (2019) implement a different approach and show that the US EPU influences the connectedness between oil and currency markets.

None of these studies focuses, however, on the recent situation generated by the COVID-19 crisis. Therefore, we fill in this gap and test the impact of coronavirus numbers and West Texas Intermediate (WTI) prices on the US EPU (we use BRENT crude for robustness purpose). As far as we know, this is the first paper addressing the impact of the COVID-19 crisis on the US policy-induced economic uncertainty.

## 3. Data and methodology

*3.1. Data*

COVID-19 data are extracted from the daily situation reports released by WHO starting with January 21, 2020. Consequently, our sample covers the period January 21, 2020 – March 09, 2020 (49 observations). The oil price data are obtained from the US Energy Information Administration (EIA), whereas the financial volatility (VIX) data comes from the Chicago Board Options Exchange (CBOE). The US EPU daily data are derived from Baker et al. (2016) and daily updated for the US on their website.[5]

Figure 1 presents the dynamics of the indicators. Figure 1(a) shows that crude oil prices record a continuous decline since the official monitoring of COVID-19, and a severe crash starting with March 6, 2020. During this period, the financial volatility and economic policy uncertainty were continuously growing. On February 17, 2020 the Chinese authorities reported 19,461 new infection cases which created a panic on the US financial markets and the transactions were closed during that day. This number is intentionally omitted from Figure 1(b) which underpin two waves of COVID-19 new infections, the first recorded in China, and the second mirroring the conditions outside China, where countries as Italy, Iran and South Korea are severely affected. The death ratio climbed to 3.5% on March 10, 2020. Figure 1(c) shows that the number of infected people overpasses the psychological barrier of 100,000, whereas the number of affected countries increases exponentially since February 24, 2020. Finally, Figure 1(d) highlights opposite movements of oil prices and death ratio associated with COVID-19 starting with the second part of February.

---
[5] Data are extracted on March 10, 2020 from http://www.policyuncertainty.com/us_daily.html.

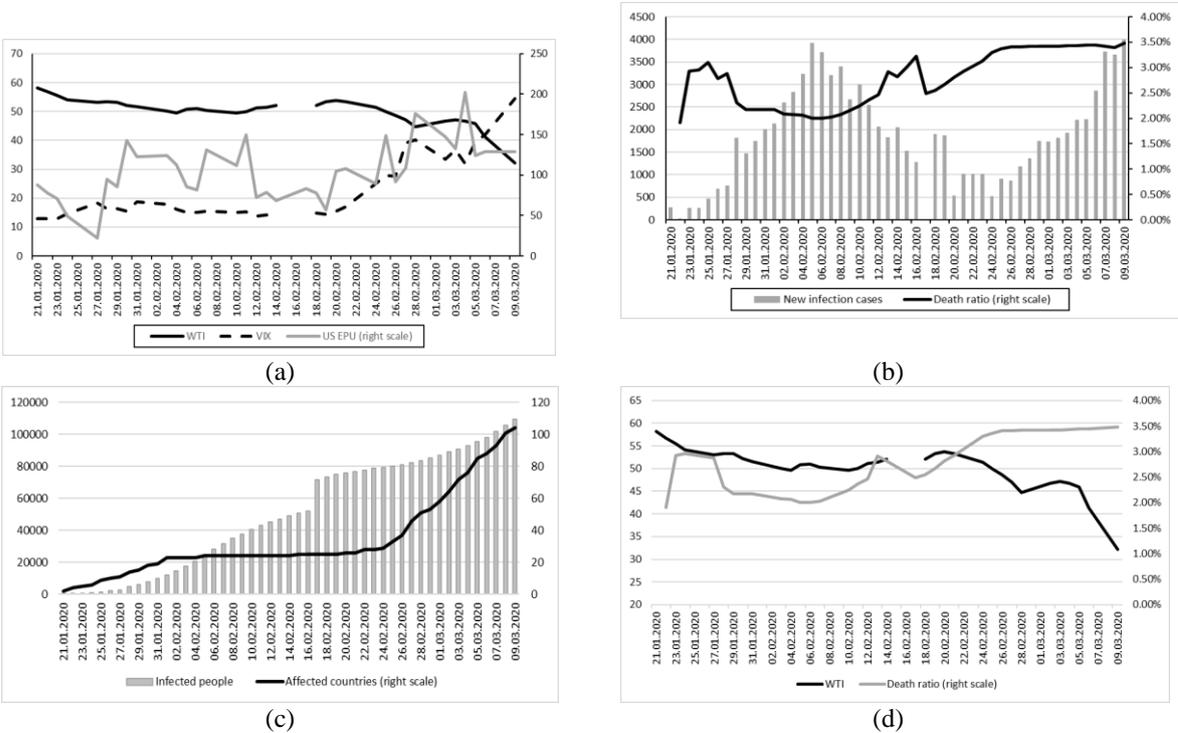

Fig. 1. Oil prices, financial volatility, economic uncertainty and coronavirus dynamics

*Sources: WHO situation reports, Chicago Board Options Exchange (CBOE), Baker et al. (2016) – daily updates*

Table 1 presents the descriptive statistics of the retained series and show a high volatility of COVID-19 new infection cases.

Table 1. Summary statistics

|          | Oil-WTI | COVID-19 | VIX   | EPU   |
|----------|---------|----------|-------|-------|
| MIN      | 32.17   | 32.00    | 12.85 | 22.33 |
| MAX      | 58.25   | 19,572   | 54.46 | 202.5 |
| MEAN     | 50.25   | 2,339    | 22.05 | 105.1 |
| ST. DEV. | 4.690   | 3,184    | 11.01 | 36.79 |

Notes: (i) COVID-19 refers to the new cases reported at global level, (ii) Oil refers to WIT prices.

The Phillips-Perron unit root test shows that our series are either I(1) or I(0). For example, the oil price and VIX series present unit roots whereas COVID-19 numbers and EPU are mean-reverting (Table 2).

Table 2. Unit root test

|  | Oil | COVID-19 | VIX | EPU |
|---|---|---|---|---|
| Level | 0.120 | -6.228*** | 2.172 | -4.785*** |
| First difference | -3.561** | -39.39*** | -5.254*** | -27.61*** |

Notes: (i) ***, ** and * means significance at 1%, 5% and 10%; (ii) the optimal lag selection is based on AIC information criterion; (iii) COVID-19 refers to the new cases reported at global level, (iv) Oil refers to WIT prices.

## 2.2. Methodology

In order to estimate the relationship between oil prices, COVID-19, VIX and EPU, we use the ARDL model proposed by Pesaran et al. (2001), who is compatible with both I(0) and I(1) series.[6]

The Pesaran et al.'s (2001) framework uses a linear transformation to integrate short-run adjustments into the long-run equilibrium, resorting to an Error Correction Model (ERM), as follows:

$$\Delta Oil_t = c + \delta_{oil} Oil_{t-1} + \delta_{COVID-19} COVID\text{-}19_t + \delta_{VIX} VIX_{t-1} + \delta_{EPU} EPU_{t-2} + \sum_{i=1}^{p} \alpha_i \Delta Oil_{t-i} +$$
$$+ \sum_{i=-1}^{p} \beta_i \Delta COVID\text{-}19_{t-i} + \sum_{i=0}^{p} \gamma_i \Delta VIX_{t-i} + \sum_{i=1}^{p} \gamma_i \Delta EPU_{t-i} + \theta ECT_{t-i} + \varepsilon_t \quad (1)$$

where: (i) c and ε is the intercept and the error term respectively, (ii) short-run terms are denoted by Δ, whereas the long-run terms are indicated with δ-terms, (iii) is the maximum number of lags (four in our case), (iv) the error correction term is denoted by ECT (θ should be negative and significant in order to validate the long-run relationship).

The optimal number of lags is selected based on the Akaike Information Criteria (AIC). The existence of a long-run relationship is validated using a F-statistic, where the null hypothesis of no cointegration is $\delta_{Oil}=\delta_{COVID-19}=\delta_{VIX}=\delta_{EPU}=0$.[7] We perform a series of post-estimation tests to check for the residual serial correlation (Breusch-Godfrey LM test)[8], the presence of ARCH effects (Engle ARCH-LM test), and normality (Jarque-Bera test).

---

[6] The WHO reports released at date "t" contains information and figures about COVID-19 reported by countries at the end of day "t-1". The press usually announces new numbers, significantly higher at date "t", which are recorded by WHO reports at date "t+1". Therefore, in our estimation we consider the impact of COVID-19 data reported at "t+1" on the oil price recorded at date "t". At the same time, the EPU index is daily updated, and might impact the oil prices with a small delay. Therefore, $EPU_{t-1}$ is considered in our general equation.

**3. Empirical results**

In the first step we check for the existence of a long-run relationship, applying the bound tests which assume a lower bound for I(0) series, and an upper bound for I(1) series, the critical values being derived from Narayan (2005). The F-statistic indicates a cointegrating relationship if the values are higher than the critical value of the upper bound. For the first two models referring to the new infection cases daily reported at global level and in China, we notice the existence of a long-run relationship between oil prices, coronavirus numbers, financial volatility and the US economic uncertainty (Table 3). However, the situation outside China can only have a short-run effect on crude oil prices.

Table 3. Bounds test results (WTI)

| Model specification | F-statistic | Critical values | | Conclusion |
| --- | --- | --- | --- | --- |
| | | Lower bound (I(0)) | Upper bound (I(1)) | |
| COVID-19 Total | 14.36 | 2.79 | 3.67 | cointegration |
| COVID-19 China | 4.004 | 2.79 | 3.67 | cointegration |
| COVID-19 Outside China | 1.607 | 2.79 | 3.67 | no cointegration |

Notes: (i) Critical values at 5% significance level.

In the second step, we estimate the three ARDL models (Table 4). For Model 1, covering the overall situation, a negative long-run connection between oil prices, coronavirus numbers, financial volatility and US economic policy uncertainty is noticed. The negative effect is stronger for the VIX, whereas COVID-19 has only a marginal effect on WTI. A similar situation is recorded in the short run, except for the influence of US EPU, which becomes insignificant.

Model 2 evidences similar findings in the long run, result which is not surprising given that in January – February 2020, the new cases of infection are dominated by those reported in China. The results shows that an increase of 100% in the number of new infections reported, lead to a decrease of oil price of 0.1%. For Model 3, there is no significant long-run relationship between our variables. However, in the short run, it appears that the effect of COVID-19 on oil prices is more important as compared to those reported in China. These results can be easily explained by the reaction of commodity and financial markets to the COVID-19 spread in Europe and the US.

---

[7] The optimal number of lags is chosen based on the Akaike Information Criteria (AIC) automatic selection criterion.
[8] Pesaran et al. (2001) show that ARDL models are free from residual correlation. Therefore, there are not endogeneity issues related to appropriate lag selection.

Table 4. Estimation of the ARDL specification (WTI)

|  | Model 1: COVID-19 Total | | Model 2: COVID-19 China | | Model 3: COVID-19 Outside China | |
| --- | --- | --- | --- | --- | --- | --- |
| Long-run equation | | | | | | |
| COVID-19$_{t+1}$ | -0.001*** | [0.000] | -0.001** | [0.000] | | |
| VIX$_t$ | -0.282*** | [0.014] | -0.200*** | [0.044] | | |
| EPU$_{t-1}$ | -0.009** | [0.004] | -0.060*** | [0.011] | | |
| c | 5.987*** | [0.310] | 6.163*** | [0.938] | | |
| Short-run equation | | | | | | |
| ΔOil$_{t-1}$ | 0.536*** | [0.108] | | | 0.770** | [0.241] |
| ΔOil$_{t-2}$ | 0.243* | [0.121] | | | -0.289 | [0.151] |
| ΔOil$_{t-3}$ | | | | | 0.498** | [0.142] |
| ΔCOVID-19$_{t+1}$ | -0.000** | [0.000] | -0.000 | [0.000] | -0.005*** | [0.000] |
| ΔCOVID-19$_t$ | -0.001*** | [0.000] | | | 0.000 | [0.001] |
| ΔCOVID-19$_{t-1}$ | 0.001*** | [0.000] | | | -0.003* | [0.001] |
| ΔCOVID-19$_{t-2}$ | 0.000** | [0.000] | | | -0.006** | [0.001] |
| ΔVIX$_t$ | -0.154*** | [0.026] | -0.142*** | [0.031] | -0.087** | [0.031] |
| ΔVIX$_{t-1}$ | | | -0.217*** | [0.031] | -0.148** | [0.043] |
| ΔVIX$_{t-2}$ | | | | | 0.120* | [0.056] |
| ΔEPU$_{t-1}$ | 0.002 | [0.002] | -0.017*** | [0.004] | -0.022*** | [0.004] |
| ΔEPU$_{t-2}$ | | | 0.017** | [0.006] | 0.016 | [0.008] |
| ΔEPU$_{t-3}$ | | | 0.021*** | [0.005] | 0.024 | [0.007] |
| ΔEPU$_{t-4}$ | | | 0.016*** | [0.004] | 0.014 | [0.004] |
| ECT$_{t-1}$ | -1.453*** | [0.151] | -0.821*** | [0.161] | -0.721** | [0.189] |
| Tests | | | | | | |
| Serial correlation | NO | | NO | | NO | |
| ARCH effects | NO | | NO | | NO | |
| Stability | YES | | YES | | YES | |

Notes: (i) ***, ** and * means significance at 1%, 5% and 10%; (ii) standard deviations are in square brackets; (iii) Breusch-Godfrey LM test for serial correlation is used; (iv) ARCH effects for conditional heteroscedasticity (with 4 lags); (v) Ramsey and CUSUM tests are used to check the stability.

## 4. Robustness analysis

The robustness results considering the BRENT prices confirm our initial findings. However, the bound tests shows that the existence of a cointegration relationship is documented at 10% significance level only (Table 5).

Table A1. Bounds test results (BRENT)

|  | F-statistic | Critical values | | Conclusion |
| --- | --- | --- | --- | --- |
| Model specification | | Lower bound (I(0)) | Upper bound (I(1)) | |
| COVID-19 Total | 3.245 | 2.79 | 3.67 | cointegration at 10% significance |
| COVID-19 China | 2.853 | 2.79 | 3.67 | inconclusive cointegration |
| COVID-19 Outside China | 3.149 | 2.79 | 3.67 | cointegration at 10% significance |

Notes: (i) Critical values at 5% significance level.

The long-run relationship is significant in this case for Model 3, also. Nevertheless, Table 6 shows that in the long run, COVID-19 has a rather reduced negative impact on crude oil price, whereas the effect is not significant for the cases reported outside China. The short-run equation reveals mixed findings.

Table 6. Estimation of the ARDL specification (BRENT)

|  | Model 1: COVID-19 Total | | Model 2: COVID-19 China | | Model 3: COVID-19 Outside China | |
| --- | --- | --- | --- | --- | --- | --- |
| Long-run equation | | | | | | |
| COVID-19$_{t+1}$ | -0.001*** | [0.000] | -0.001* | [0.001] | 0.003 | [0.001] |
| VIX$_t$ | -0.213*** | [0.044] | -0.391* | [0.242] | -0.495** | [0.181] |
| EPU$_{t-1}$ | -0.041** | [0.014] | -0.056 | [0.081] | -0.120*** | [0.018] |
| c | 6.663*** | [0.705] | 7.124*** | [0.933] | 7.546*** | [0.363] |
| Short-run equation | | | | | | |
| ΔOil$_{t-1}$ | 0.560** | [0.239] | -0.351 | [0.226] | 0.094** | [0.571] |
| ΔOil$_{t-2}$ | | | -0.407 | [0.269] | | |
| ΔOil$_{t-3}$ | | | -0.183 | [0.238] | | |
| ΔCOVID-19$_{t+1}$ | -0.000 | [0.000] | -0.000 | [0.000] | -0.000 | [0.000] |
| ΔCOVID-19$_t$ | -0.001*** | [0.000] | -0.000 | [0.000] | 0.000** | [0.001] |
| ΔCOVID-19$_{t-1}$ | 0.001 | [0.000] | -0.000 | [0.000] | 0.008*** | [0.001] |
| ΔCOVID-19$_{t-2}$ | 0.000** | [0.000] | | | -0.003 | [0.002] |
| ΔVIX$_t$ | -0.076 | [0.057] | -0.173* | [0.079] | -0.060 | [0.053] |
| ΔVIX$_{t-1}$ | -0.113 | [0.066] | -0.268** | [0.091] | -0.068 | [0.067] |
| ΔVIX$_{t-2}$ | 0.088 | [0.076] | -0.022 | [0.114] | 0.160* | [0.075] |
| ΔVIX$_{t-3}$ | -0.158* | [0.087] | 0.015 | [0.101] | | |
| ΔEPU$_{t-1}$ | -0.015** | [0.006] | -0.014 | [0.009] | -0.038*** | [0.007] |
| ΔEPU$_{t-2}$ | 0.028* | [0.013] | -0.012 | [0.011] | 0.041** | [0.014] |
| ΔEPU$_{t-3}$ | 0.030*** | [0.008] | | | 0.055*** | [0.013] |
| ΔEPU$_{t-4}$ | | | | | 0.038*** | [0.009] |
| ECT$_{t-1}$ | -1.443*** | [0.292] | -0.401** | [0.158] | -0.876*** | [0.180] |
| Tests | | | | | | |
| Serial correlation | NO | | NO | | NO | |
| ARCH effects | NO | | NO | | NO | |
| Stability | YES | | YES | | YES | |

Notes: (i) ***, ** and * means significance at 1%, 5% and 10%; (ii) standard deviations are in square brackets; (iii) Breusch-Godfrey LM test for serial correlation is used; (iv) ARCH effects for conditional heteroscedasticity (with 4 lags); (v) Ramsey and CUSUM tests are used to check the stability.

All in all, we notice that even if the direct negative impact of COVID-19 new infection cases on crude oil prices is rather limited, the indirect effect manifested through the volatility of financial markets cannot be neglected. Given the speed of propagation of this pandemic virus, if the world governments are not proactive, promptly implementing the required measures to isolate the suspected cases of COVID-19, the global economy risk to be paralyzed in few weeks.

## 5. Conclusions

The new coronavirus has generated noteworthy shock waves on financial markets, but also on commodity prices, including oil. Oil prices recorded the hardest cut after 1991, which help, for the moment, the economy of oil-importing countries severely affected by the coronavirus crisis. However, the crash of oil prices clearly shows that an economic downturn cannot be avoided.

In this context, the purpose of our paper was to see how the COVID-19 numbers, in terms of daily announcements of new infection cases, influenced the international oil prices. Our ARDL estimation documented a negative and significant impact of the coronavirus crisis, but relatively small as compared to the effect of financial volatility and economic policy uncertainty on oil prices. The COVID-19 impact on oil prices seems to be rather indirect, affecting first the financial markets volatility (Albulescu, 2020).

Even if the figures reported outside China seem to have, for the moment, no significant effect on oil prices in the long run, the exponential increase of new infection cases risks to block the world economy and to freeze oil prices at a low level for a long period. The amplitude of the economic contraction will be correlated with the coronavirus persistence. Although China seems to gain the fight against COVID-19, the virus exponentially propagates in Europe and the US. Consequently, a strong coordinated worldwide reaction is required, including economic measures to prevent a severe economic downturn. Central banks have already started to cut the interest rates, but this measure should be followed by appropriated fiscal facilities.

**Acknowledgements**

This work was supported by a Grant of the Romanian National Authority for Scientific Research and Innovation, CNCS–UEFISCDI, Project Number PN-III-P1-1.1-TE-2019-0436.